\documentstyle[12pt,epsfig]{article}
\input{epsf}
\newcommand{\be}{\begin{equation}}
\newcommand{\ee}{\end{equation}}

\begin{document}

\title{Broad resonances as locking states and the problem of
confinement }
\author{V.V.Anisovich$^{a1}$, D.V.Bugg$^{b2}$ and 
A.V.Sarantsev$^{a3}$}
\date{} 
\maketitle

\begin{abstract}
We discuss the mechanism of broad state formation when one of several
overlapping resonances accumulates the widths of others. This mechanism
may be important in the mixing of  quark states with exotic ones.
The broad resonance acts as a locking state for other states which mix
with it; we consider the role of the broad state in
the formation of the confinement barrier.
\end{abstract}

\vspace{4cm}

{\parindent=0cm
\line(1,0){290}
\footnotesize

a) {\it St.Petersburg Nuclear Physics Institute, Gatchina, 
188350 Russia}\\
b) {\it Queen Mary and Westfield College, London E14NS, UK}\\  
1) anisovic@lnpi.spb.su,  \ anisovic@thd.pnpi.spb.ru\\
2) bugg@v2.rl.ac.uk   \\
3) vsv@hep486.pnpi.spb.ru, \  andsar@v2.rl.ac.uk }

\newpage
In a series of papers, we have analysed the
$\pi \pi ~(IJ^{PC}=00^{++})$-wave \cite{bsz,km1900,n/d}.  The first
paper uses the T-matrix and the second the
$K$-matrix approach for the mass region below 1900 MeV.  The third
reconstructs the propagator matrix for the scalar/isoscalar states.
These analyses indicate the existence
of a broad state $f_0(1530{+90 \atop-250})$ with half-\-width
$\Gamma /2\sim 500$ MeV.
According to \cite{km1900,n/d}, this resonance carries roughly
50\% of the gluonium component, being a descendant of the
lightest pure glueball state.
This should not come as a surprise.
Stationary $q\bar q$ states are orthogonal to one another,
though loop diagrams
involving their decays cause some mixing.
Via this mixing, the
widths of overlapping resonances may be accumulated by one of them,
while the others transform into rather narrow states.
The glueball may mix freely with all $q\bar q$ states:
according to the
roles of 1/N-expansion \cite{1/n},
$q\bar q$/gluonium mixing is not suppressed
\cite {n/d,ufn}. That is the reason for the large
gluonium component in the broad resonance.

Capturing the large portion of the widths of the neighbouring states,
the broad resonance plays another function as well: it
plays the role of a locking state for the other resonances. All other
states, being narrow ones, are now confined to
the small-$r$ region by the locking state.
Experiment gives us evidence that the locking state
is mainly located at comparatively large distances.

The same effect may take place for other exotic hadrons;
by this we mean glueballs and hybrids with different quantum numbers.
Therefore, we may expect the existence of the locking states in all
partial waves and in  different mass regions.
It makes it possible that the broad states play an important role
in the confinement phenomenon, creating a dynamical barrier for
their counterparts.

\section{Effect of accumulation of widths in the \newline $K$-matrix
approach}
To see the mixing of non-stable states in a pure form, let us consider
an example with three resonances decaying into the same channel. In the
$K$-matrix approach, the amplitude we consider is written in the form:
\be
A = K(1-i\rho K)^{-1} \; , \qquad
K = g^2\sum_{a=1,2,3}\
(M^2_a-s)^{-1}\; .
\ee
Here, for purposes of illustration, we take $g^2$ to be the  same for
all 3 resonances, and make the approximations that:  (i)~the phase space
factor $\rho$ is constant, and (ii)~$M^2_1=m^2-\delta$, $M^2_2=m^2$,
$M^2_3=m^2+\delta$.
Fig.1 shows the location of the poles in the
complex-$\sqrt s$ plane $(\sqrt s=M-i\Gamma/2)$ as the coupling $g$
increases.
At large $g$, which corresponds to a strong overlapping of the
resonances, one resonance accumulates the widths of the others while
two counterparts of the broad state become nearly stable.

Figs. 2a and 2b show the actual situation for the $00^{++}$-wave 
\cite{km1900}. The figures show the  movement of the poles  as
the parameters $\xi$ is varied.
There we replace the $K$-matrix elements for solution I from
\cite{km1900},
\be
K_{ab}\ =\ \sum_\alpha \frac{g^{(\alpha)}_a g^{(\alpha)}_b}{
M^2_\alpha-s}+f_{ab}\ \to\ \sum\xi\;
\frac{g^{(\alpha)}_ag^{(\alpha)}_b}{M^2_\alpha-s}+f_{ab}\; ,
\ee
varying $\xi$ in the interval $0<\xi\le 1.$
The case $\xi =1$ shows the
pole positions for the states $f_0(980)$, $f_0(1300)$, $f_0(1500)$,
$f_0(1530{+90 \atop-250})$ and $f_0(1750)$.
As $\xi \to 0$ the decay
processes and the corresponding mixing are switched off;  the
positions of the poles show the masses of the bare states:
$f^{bare}_0(720\pm10)$, $f^{bare}_0(1230\pm50)$,
$f_0^{bare}(1260\pm30)$, $f^{bare}_0(1600\pm50)$ and
$f^{bare}_0(1810\pm30)$.
The broad resonance $f_0(1530
{+90\atop-250})$, which is seen experimentally to contain a large
component due to the scalar gluonium, has
accumulated the widths of the neighbouring $q\bar q$-states.  This
prevents them from decaying quickly.

\section{The broad resonance as a locking state}

We now discuss the role of the broad resonance in the formation of the
hadron spectrum.
Let us consider as a guide the meson spectrum in the
standard quark model with decay processes switched off (Fig.3a).
Here we have a set of stable levels. If the decay processes are 
incorporated
for the highly excited states, it would be naive to think that the
decay processes result only in broadening of levels.
Due to processes
$bare\; state\to real\; mesons$, the resonances mix and one of them
may transform into the very broad state.
This creates a trap for the states with which it overlaps.
Fig.3a shows the conventional potential model.
In Fig.3b we show instead a confinement barrier through which 
states may decay.
The broad locking state appears outside the barrier.
The comparatively narrow locked  states lie inside the confinement 
well. Therefore, the broad
resonance plays the role of a dynamical barrier which prevents decay of
states which are left in the small-$r$ region.
This is the familiar
phenomenon that an absorption process acts as a reflecting barrier.

From the point of view of the analysis of the $(IJ^{PC}=00^{++})$-wave
performed in \cite{n/d}, the scalar glueball state lies "by
chance" among $n\bar n$-dominant states
(we denote $n\bar n =(u\bar u+d\bar d)/ \sqrt{2}$).
It mixes with $1^3P_0(n\bar n\;rich)$ and
$2^3P_0(n\bar n\;rich)$ states and transforms into the broad resonance.
So the lightest scalar glueball plays "accidentally" the role of creating
the locking state for the $00^{++}$-wave in the mass region
1100--1600 MeV.

It seems quite reasonable to expect that the same situation
may happen with other exotic states: glueballs and hybrids.
In this
case the existence of comparatively narrow resonances (the locked
states) and a very broad one (the locking one) is likely to be a
signature of an exotic state.

A second example is now known for $J^P = 0^{-+} $ \cite{Zerominus},
where a very broad $0^{-+}$
state around 1800--2100 MeV has been identified with decays which are
approximately flavour-blind.
That paper also advances arguments for a broad $2^{++}$ resonance
at 2000--2400 MeV,
though there is not yet direct experimental proof of such a broad
state.  

\section{Locking states and confinement}

Let us consider the diffractive dissociation of mesons in high energy
collisions: in this process, the mechanism of quark deconfinement
reveals itself.
The typical process of mesonic diffractive
dissociation is shown in Fig.4a.
The pomeron transfers a momentum to
the interacting quark, so that the kicked quark can leave the
confinement trap creating a new quark-antiquark pair, thus transforming
into a white meson.
The diffractive dissociation cross section is
determined by cutting the diagram of Fig.4b, where the pomerons play
the role of momentum/energy injectors into the self-energy meson
diagram.
We suppose that the contribution
of the loop diagrams of Fig.4c (or, more generally,
diagrams of Fig.4d) at $ r\sim R_{confinement}$ is directly related to
the confinement process.
For the highly excited states, the
diagrams of Figs.4c-4d are just those which are responsible for the
formation of the broad state. (Only the planar diagrams which  give the
 leading contributons in the $1/N$-expansion \cite{1/n} are
 drawn in Fig.4).

Two different $r$-regions with different physics contribute to the
diagrams of Figs. 4c-4d: the small-$r$ region, $r\leq  
R_{confinement}$,
and the large-$r$ one, $r\geq  R_{confinement}$.
Within the region $r\leq  R_{confinement}$,
QCD-based states are formed ($q\bar q$-levels and exotic mesons like
glueballs and hybrids).
The interaction in the large-$r$ region plays an important role
in the creation of the broad locking states.

There is direct experimental evidence that the broad state is 
associated with
large $r$.  In the GAMS data on $\pi ^- \pi ^+ \to \pi ^0 \pi ^0$ 
\cite{GAMS},
the broad state is clearly visible for $|t| < 0.2$ GeV$^2$, but 
disappears at
large $|t|$, leaving peaks due to narrower states clearly visible.

The most convenient language for a description of the large-$r$ 
interaction
is the language of hadron states.
The block inside the self-energy
diagram of Fig.4d represents a set of diagrams of the Bethe-\-Salpeter
type for two-\-meson scattering via meson exchange.
It offers hope that the physics of the locking state may be
calculated quantitatively by use of the Bethe-\-Salpeter equations 
with
mesonic degrees of freedom only (see refs. \cite{BZ}, 
\cite{Julich} and references therein).
Alternatively,  if the small-$r$ interaction is important for this 
physics,
it may be possible to use hybrid models where both mesonic
and quark/gluon degrees of
freedom are taken into account; examples of this type of consideration
are given by \cite{as-conf}, \cite{hyb-b}.

\section{Conclusion}

In the deconfinement of quarks of an exited $q\bar q$-level, there are
two stages:

(i) An inevitable creation of new quark-antiquark pairs which result
in production of  white hadrons. This stage is the subject
of QCD and  is beyond our present discussion.

(ii) The outflow of the created white hadrons and their mixing
results in the production of a very broad state. The broad resonance
locks other $q\bar q$-levels into the small-$r$ region,
thus playing the role of a dynamical barrier;
this is the reason for calling the broad resonance  a locking state.

The bare states are  the subjects of the quark/gluon classification.
Exotic hadrons like glueballs and 
hybrids mix readily with $q\bar q$ states and
are good candidates to generate locking states in all waves.

The question, which is not answered yet, is if the broad states are
created by the large-$r$  component of the forces alone (i.e. 
$t$-channel
exchanges). An
investigation of realistic Bethe-Salpeter type equations with and
without inclusion of the small-$r$ component of the forces appears 
to be the
way to make quantitative progress in examining  this point.

Rich physics is hidden in the broad states, and an investigation of
them is an important and unavoidable step in understanding the 
spectroscopy of
highly exited states and their confinement.
\vskip 0.5cm
{\bf Acknowledgements}\\
This work was supported by INTAS-RFBR grant 95-0267.

\newpage
\begin{figure}
\begin{center}
\epsfig{file=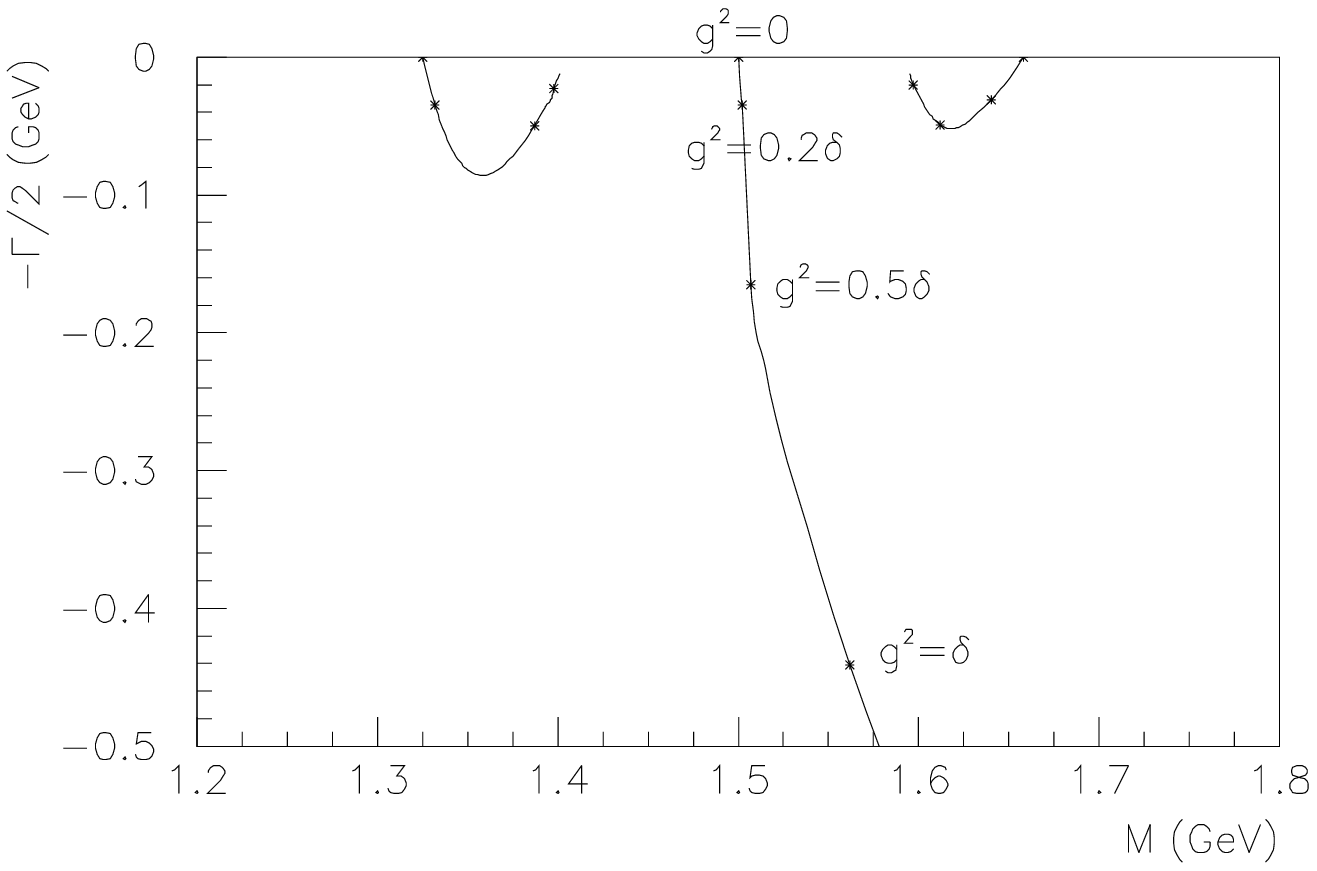,width=13.5cm}\\
Fig.1: Position of the poles of the amplitude of Eq.(1) 
in the complex-$\sqrt s$ plane ($\sqrt s=M-i\Gamma/2$) with increase
of $g^2$; in this example 
$m=1.5$ GeV, $\delta=0.5$ GeV$^2$ and the phase space factor 
is fixed: $\rho = 1$.
\end{center}
\end{figure}
\newpage
\begin{figure}
\begin{center}
\epsfig{file=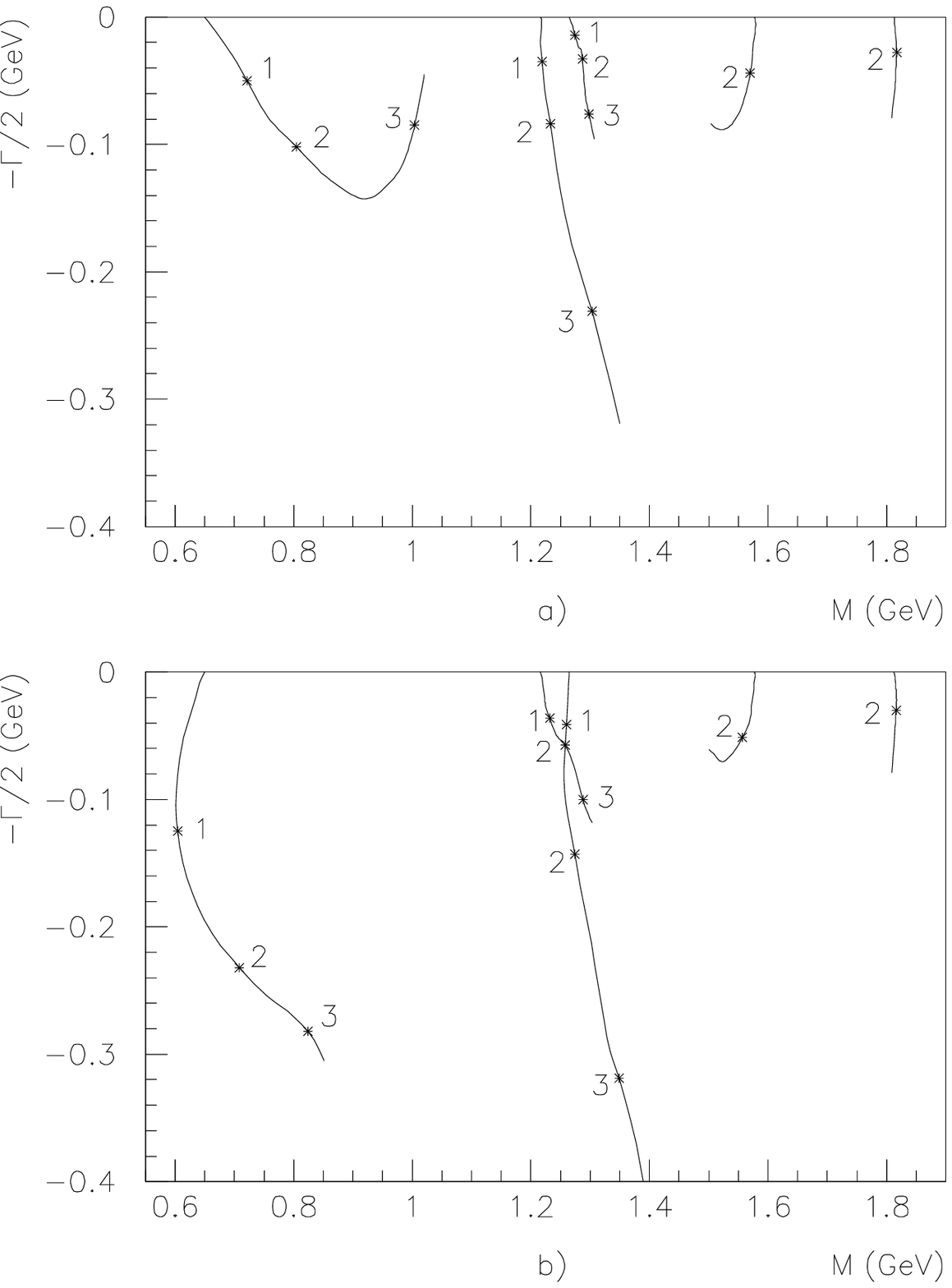,width=13.5cm}\\
Fig.2 Positions of the poles in the 
complex-$\sqrt s$ plane ($\sqrt s=M=-\Gamma/2$)
with varying  $\xi$ on the sheet under $\pi\pi$ cut (a) and
on the sheet under $K\bar K$ cut (b) from 
$K$-matrix representation of the $(IJ^{PC}=00^{++})$-amplitude 
obtained in \cite{km1900}.
\end{center}
\end{figure}
\newpage
\begin{figure}
\begin{center}
\epsfig{file=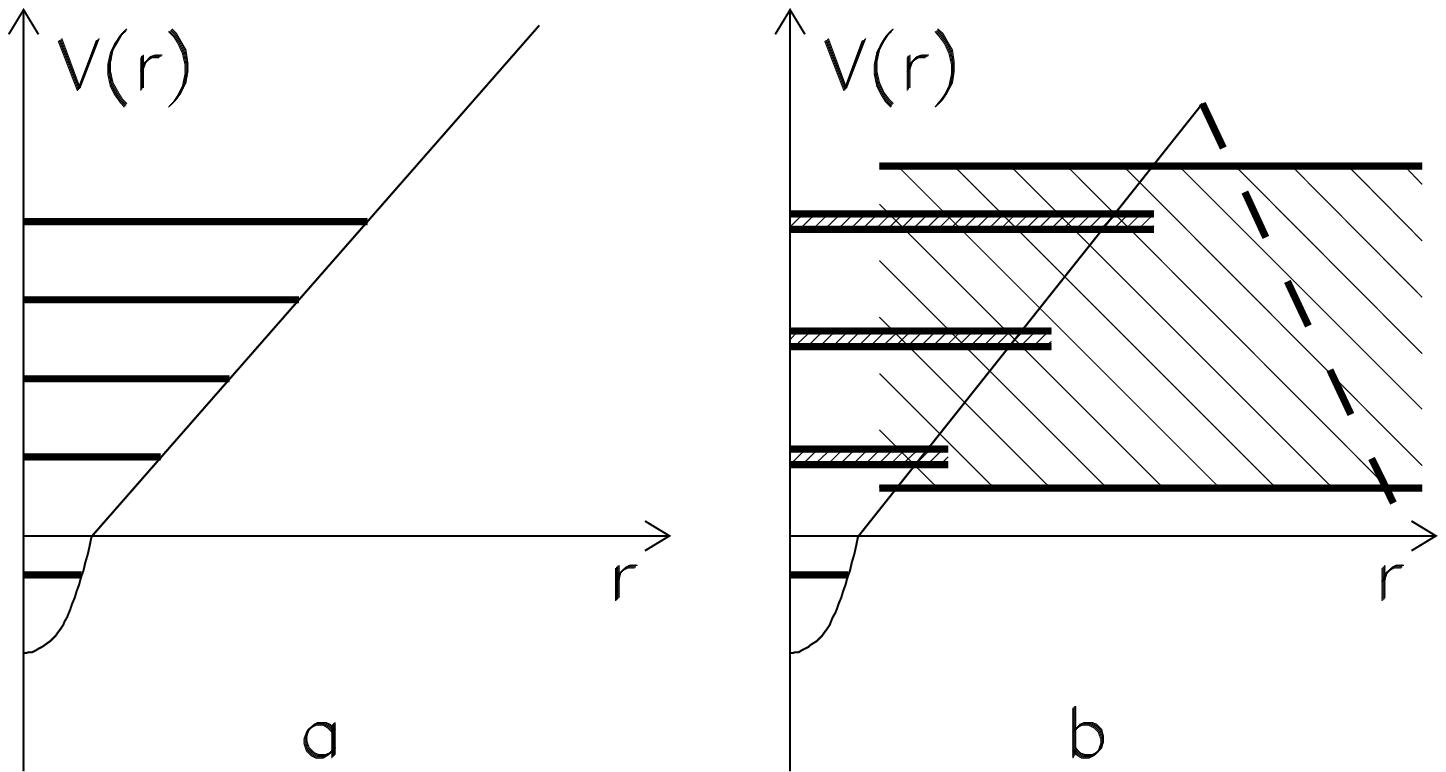,width=14cm}\\
~\\
~\\
Fig.3 Conventional pictures for the potential model levels
without 
decay processes taken into account (a) and with them (b).
\end{center}
\end{figure}
\newpage
\begin{figure}
\begin{center}
\epsfig{file=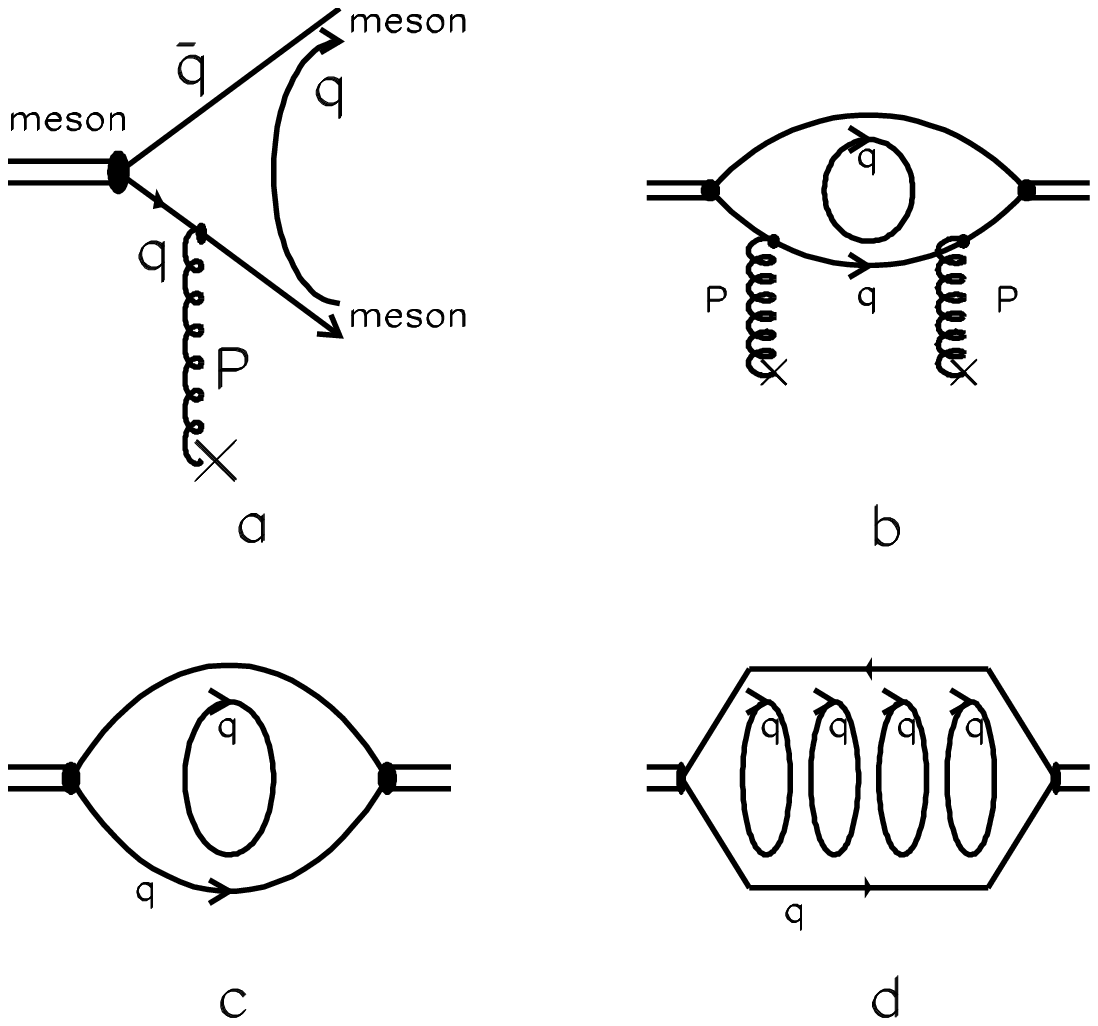,width=14cm}\\
Fig.4. a) Diffraction dissociation (DD) process: quarks leave the
confinement trap creating at comparatively large distances a new
quark-\-anti-quark pair. b)~Diagram responsible for $DD$-cross section;
c,d)~Self-energy diagrams for large-$r$ contributions related to
the creation of the locking state.
\end{center}
\end{figure}

\end{document}